# Recent advancements in the study of intrinsic magnetic topological insulators and magnetic Weyl semimetals

Wei Ning and Zhiqiang Mao*

Department of Physics, The Pennsylvania State University, University Park, Pennsylvania 16802, USA

**Abstract**

The studies of topological insulators and topological semimetals have been at frontiers of condensed matter physics and material science. Both classes of materials are characterized by robust surface states created by the topology of the bulk band structures and exhibit exotic transport properties. When magnetism is present in topological materials and breaks the time-reversal symmetry, more exotic quantum phenomena can be generated, e.g. quantum anomalous Hall effect, axion insulator, large intrinsic anomalous Hall effect, etc. In this research update, we briefly summarize the recent research progresses in magnetic topological materials, including intrinsic magnetic topological insulators and magnetic Weyl semimetals.

*zim1@psu.edu

I.   Introduction

Magnetic topological materials, including magnetic topological insulators (TI) and magnetic topological semimetals, have attracted broad interests. Magnetic TIs can be achieved in three different ways: magnetic doping in a TI [1,2], proximity of a TI to a ferromagnetic (FM) or an antiferromagnetic (AFM) insulator [3,4,5], or creating intrinsic FM or AFM order in a TI [6]. The spontaneous magnetization induced in magnetic TIs interacts with topological surface states and opens a gap at the surface Dirac point, which can generate a new topological quantum state - quantum anomalous Hall insulator (QAHI), when chemical potential is tuned to an appropriate value in thin film samples. Since QAHI features spin-polarized chiral edge state, which can support dissipationless current, it carries great promise for applications in future energy saving electronics.  QAHI was first realized in thin TI films of Cr- and/or V-doped $(Bi,Sb)_2Te_3$ [1,2]. This pioneering work has generated a great deal of interest and several review articles [7,8,9,10,11] on this topic have been published. In this research update, we will focus on reviewing recent studies on intrinsic magnetic TI $MnBi_2Te_4$ and its related materials.

In magnetic topological semimetals, the interplay between magnetism and non-trivial band topology can also generate new exotic quantum states.  One remarkable example is time reversal symmetry (TRS) breaking Weyl semimetal (WSM) state in which linearly dispersed, spin-split bands cross at discrete momentum points, thus resulting in Weyl nodes. Low energy excitations near Weyl nodes behave as chiral Weyl fermions. Weyl nodes always come in pairs with opposite chirality and they can be understood as source and drain of Berry curvature in momentum space. When the Weyl nodes are at or close to the Fermi level, net Berry curvature can be present due to broken TRS, which can give rise to new exotic quantum phenomena such as large intrinsic anomalous Hall effect (AHE) [12] and anomalous Nernst effect [13]. Like non-magnetic WSMs, magnetic WSMs are also characterized by topological surface states, i.e. surface Fermi arcs [14]. Theory has predicted TRS breaking WSMs can also evolve into QAHI when the dimensionality is reduced from 3D to 2D [15]. Experimentally, several materials, including $Co_3Sn_2S_2$ [16,17,18,19], $Co_2MnGa$ [20,21], $Co_2MnAl$ [22], $Mn_3Sn$ [23], GdPtBi [24] and $YbMnBi_2$ [25] have been reported to be TRS breaking WSM states. In this research review, we will also give a brief overview on the studies of these materials.

## II. Intrinsic magnetic topological insulator $MnBi_2Te_4$ and its related materials

Although quantum anomalous Hall effect (QAHE) has been seen in thin TI films of Cr- and V-doped $(Bi,Sb)_2Te_3$ [1,2], the 'critical temperature' required is below ~2 K, severely constraining the exploration of fundamental physics and technological applications. Inhomogeneous surface gap induced by randomly distributed magnetic dopants is believed to be the origin of low-temperature requirement for observing QAHE [26,27]. High-temperature QAHE has been predicted to occur in thin films of intrinsic FM or AFM TI materials [15,28]. Nevertheless, despite considerable theoretical and experimental efforts, there has been little progress until the recent discovery of an intrinsic AFM TI $MnBi_2Te_4$ [29,30,31]. $MnBi_2Te_4$ is a layered ternary tetradymite compound; it crystallizes in a rhombohedral structure (space group R-3m), built of the stacking of Te-Bi-Te-Mn-Te-Bi-Te septuple layers (SLs) (Fig. 1a). SLs are coupled through van der Waals bonding.

The single crystals of $MnBi_2Te_4$ can be grown either from the melt with stoichiometric composition [31, 32] or using the flux method with excessive $Bi_2Te_3$ serving as flux [33]. Since $MnBi_2Te_4$ is metastable [34], its single crystals can be obtained only through quenching at a temperature close to 590°C. For the melt growth, the stoichiometric mixture first needs to be heated to a high temperature (700-1000°C), then slowly cooled down to a temperature close to 590°, finally followed by annealing and quenching at this temperature[31,32]. For the flux growth, prolonged slow cooling (~2 weeks) from ~600°C to ~590°C is necessary, and the excessive flux is separated through centrifuging [33].

$MnBi_2Te_4$ enables combination of intrinsic antiferromagnetism with nontrivial band topology, thus giving rise to an intrinsic AFM TI [29,30,31]. Its antiferromagnetism is produced by the Mn-sub-lattice, while its nontrivial band topology is formed by inverted Bi and Te $p_z$ bands at the Γ point due to strong spin-orbit coupling (SOC). Its AFM state shows an *A*-type AFM order ($T_N$ = 25K) [31,33,35], characterized by Mn FM layers stacked antiferromagnetically along the *c*-axis and the ordered magnetic moments are aligned along the *c*-axis [33]. A large spin gap as well as magnetic frustration due to large next-nearest neighbor AFM exchange have also been probed in recent inelastic neutron scattering experiments on $MnBi_2Te_4$ [36]. On the (001) surface, a large gap (~88 meV [31]) is opened at the surface Dirac node due to the breaking of the $S=\theta T_{1/2}$ symmetry ($\theta$ and $T_{1/2}$ represent the time reversal and primitive translation symmetry respectively) (Fig. 1b). Such a surface gap was probed in ARPES measurements on single crystal samples first by Otrokov et al. [31] (Fig. 1d) and subsequently by several other groups [37,38]. However, there have also been reports on ARPES experiments which [39,40,41,42] show the surface Dirac cone state is gapless either in the paramagnetic or the AFM state (Fig. 1e).

$MnBi_2Te_4$ offers an ideal platform to realize new exotic topological quantum states. Theory predicts it can host not only high-temperature QAHE and axion insulator with topological magnetoelectric effect in thin film samples [29,30,31, 43], but also an ideal Weyl semimetal state with one pair of Weyl nodes near the Fermi level in its bulk FM phase driven by external magnetic fields or strain (Fig. 1c) [29,30]. Moreover, chiral Majorana mode is also predicted to be accessible via interaction between MnBi$_2$Te$_4$ and a s-wave superconductor [44]. Recently, remarkable progresses have been made toward realizing these predicted quantum states [45,46].

Deng *et al.* [45] firstly reported the observation of quantized Hall resistance of $h/e^2$ ($h$ is the plank constant and $e$ is the elemental charge) in atomically thin MnBi$_2$Te$_4$ flakes with odd number of SLs (i.e. 5 SLs) under zero magnetic fields (Fig.2a). Such quantized Hall resistance is accompanied by zero longitudinal resistance, which is typical behavior of QAHI. Contrasted with the conventional QHE in 2D electron gas, QAHE in MnBi$_2$Te$_4$ does not originate from quantized Landau levels. When MnBi$_2$Te$_4$ is exfoliated to flakes with even number of SLs (e.g. 6 SLs), it is found to exhibit axion insulator behavior at zero magnetic field, characterized by large longitudinal resistance and zero Hall resistance [46] (Fig.2b); moderate magnetic fields can drive the axion insulator to the Chern insulator with quantized Hall resistance of $h/e^2$ (Fig. 2c).

Furthermore, another interesting result observed in MnBi$_2$Te$_4$ atomic crystals is the high Chen number QAHE ($C$=2) (Fig. 2d and Fig.2e) [47]. High Chern number QAHE is of interest in view of applications, since a high Chern number could enable the chiral edge states to carry larger current. Quantum confinement effect induced by dimensionality reduction should play an important role in realizing the QAHE and axion insulator in the 2D thin layers of MnBi$_2$Te$_4$. In few-layer thin films, the surface states should dominate its longitudinal transport properties and the two surfaces (top and bottom) display half-integer Hall conductance of opposite (axion insulator) or identical sign (QAHE). However, for thick flakes, interlayer coupling affects its transport properties, leading to very different transport behavior from few-layer thin flakes [47].

This may explain why thinner MnBi$_2$Te$_4$ flakes show QAHE or axion insulator, but thicker flakes (9 or 10 SLs) behave as a high Chern number insulator.

Additionally, experimental studies on MnBi$_2$Te$_4$ and Mn(Bi,Sb)$_2$Te$_4$ bulk single crystals have also revealed many other interesting properties. First, Lee et al. [32] found MnBi$_2$Te$_4$ undergoes two magnetic transitions upon increasing magnetic field (parallel to the c-axis), i.e. the spin-flop transition from an AFM to a canted antiferromagnetic (CAFM) state at H$_{c1}$ (~3.6 T) and the CAFM-to-FM transition at H$_{c2}$ (~7.7T). The CAFM state shows intrinsic AHE due to the non-collinear spin structure [32]. Second, both magnetism and carrier density in MnBi$_2$Te$_4$ are found to be tunable by Sb substitution for Bi. Single crystals of the Mn(Bi$_{1-x}$Sb$_x$)$_2$Te$_4$ alloy series with $0 \leq x \leq 1$ can be made using a similar flux growth method used for growing MnBi$_2$Te$_4$ [48,49,50]. Both H$_{c1}$ and H$_{c2}$ are suppressed by Sb substitution for Bi and merge as $x$ approaches 1 [49]. As $x$ is equal or close to 1, the system involves strong competition between FM and AFM phases [49]. Both FM and AFM phases have been synthesized for MnSb$_2$Te$_4$ [49, 51] and MnSb$_{1.8}$Bi$_{0.2}$Te$_4$ [50, 52]. FM MnSb$_2$Te$_4$ is predicted to host either an ideal type-II Weyl semimetal phase [51], or the simplest type-I Weyl semimetal with only one pair of Weyl nodes on the three-fold rotational axis under strain tuning [53], while FM MnSb$_{1.8}$Bi$_{0.2}$Te$_4$ has been reported to show unusual AHE [52]. In the AFM Mn(Bi$_{1-x}$Sb$_x$)$_2$Te$_4$ series, the carrier density can effectively been tuned by changing Sb concentration, down to a minimum near $x = 0.3$ where the carrier type also changes from electron to hole [48,49,50]. Such a critical composition could favor the observation of QAHE. The realization of QAHE state generally requires the chemical potential to be inside the gap to achieve a bulk insulating state. For pristine MnBi$_2$Te$_4$, a relatively large gate voltage is required to tune it to such a state since as-grown MnBi$_2$Te$_4$ crystals are always heavily electron doped [45] . If crystals of Mn(Bi$_{1-x}$Sb$_x$)$_2$Te$_4$ ( $x$ ~0.26) was used in the devices, QAHE can probably be seen at much smaller gate voltages. Moreover, Sb substitution for Bi increases the surface gap [54], which might increase the observation temperature of QAHE. Additionally, Lee et al.[50] recently reported the predicted ideal Weyl state can be achieved in the CAFM and FM phases of the samples with minimal carrier density. This is revealed by a magnetic-field induced electronic phase transition at the AFM-to-FM phase boundary, a large intrinsic anomalous Hall effect (see Fig. 1f), a non-trivial π Berry phase of the cyclotron orbit and a large positive magnetoresistance in the FM phase [50].

In addition to $MnBi_2Te_4$, several van der Waals materials relevant to $MnBi_2Te_4$, including $MnBi_4Te_7$, $MnBi_6Te_{10}$ and $MnBi_8Te_{13}$, are also recently reported to be intrinsic AFM/FM TI/axion insulator [55,56,57,58,59,60,61,62,63]. These materials belong to the same family, which can be expressed as $(MnBi_2Te_4)(Bi_2Te_3)_m$ with $m$ = 1, 2, 3, …Their common structural characteristic is the alternating stacking of $[MnBi_2Te_4]$-SLs and $[Bi_2Te_3]$ quintuple layers (QLs). The main difference between the $m$=1,2,3 members is the number of QLs ($m$) sandwiched between SLs; $m$=1 for $MnBi_4Te_7$, $m$=2 for $MnBi_6Te_{10}$, and $m$=3 for $MnBi_8Te_{13}$. The magnetic properties of these materials depend on $m$. With increasing $m$, the interlayer AFM coupling becomes weak due to increased separation distance between Mn magnetic layers. Although $MnBi_4Te_7$ and $MnBi_6Te_{10}$ remain AFM, their Neel temperature decrease to 13.0 K and 11.0 K respectively [55,57]. $MnBi_8Te_{13}$, however, becomes FM with the Curie temperature of 10.5K [60], indicating interlayer magnetic coupling involves competition between AFM and FM and larger separation distance favors FM coupling. We note FM $MnBi_6Te_{10}$ with $T_C$ =12K as well as a AFM-to-FM transition in $MnBi_4Te_7$ were also reported [59, 64], suggesting the Gibbs energy difference between AFM and FM phases for these compositions is very small. Band structure calculations and ARPES studies [55,58,59,60,61] have shown all these materials host topological phases: MnBi$_4$Te$_7$ is an intrinsic AFM TI [55], whereas MnBi$_6$Te$_{10}$ is either an AFM axion insulator [61] or a FM TI [59]. MnBi$_8$Te$_{13}$ is reported to be an intrinsic FM axion insulator [60]. Another common property of these materials is that they all show large magnetic hysteresis and low spin-flip transition fields. Therefore, they offer a new promising platform to explore novel topological quantum states, including QAHE and axion insulator at high temperatures.

## III.  Magnetic Weyl semimetals

Three-dimensional (3D) Dirac semimetals, which were first theoretically predicted and experimentally verified in Na$_3$Bi [65,66,67] and Cd$_3$As$_2$ [68,69,70], can be viewed as a 3D graphene. A Dirac semimetal can transform into a WSM by breaking either the time-reversal (TRS) or inversion symmetry. Inversion symmetry broken WSMs were first discovered in non-magnetic TaAs-class materials [71,72,73,74,75]. The TRS-breaking WSM were initially predicted in Re$_2$Ir$_2$O$_7$ (Re=rare earth) [14], HgCr$_2$Se$_4$ [15] and recently demonstrated in several magnetic materials systems, including Co$_3$Sn$_2$S$_2$ [16,17,18,19], Co$_2$MnGa [20, 76], Co$_2$MnAl [22], Mn$_3$Sn/Mn$_3$Ge [23, 77], GdPtBi [24] and YbMnBi$_2$ [25]. In this section, we will review the recent research progress in the study of these magnetic WSMs.

1）**Ferromagnetic WSMs**

**1a. Kagome-lattice WSM Co$_3$Sn$_2$S$_2$:** The kagome lattice is known to host exotic quantum states such as spin liquid [78]. Recent studies show a layered FM compound Co$_3$Sn$_2$S$_2$ with Kagome-lattice (space group, R-3m) hosts a TRS breaking WSM state[16,17,18,19]. The magnetic properties of this material originate from the kagome-lattice of cobalt, whose magnetic moments order ferromagnetically and are oriented along the out-of-plane direction in the ground state (Fig. 3a) [16]. Recent μSR experiments showed such an out-of-plane FM order sustains up to 90K, and then evolves into a mixed phase of the out-of-plane FM and the in-plane AFM order in the 90-172K range, and finally to a mixed phase of paramagnetic and FM in the 172-175K range [79]. The $C_{3v}$-rotation and inversion symmetries of this material generates a total of six nodal rings without considering SOC. When SOC is considered, the linear crossing points of nodal rings split into three pairs of Weyl modes as shown in Fig. 3d. These Weyl nodes are only about 60 meV above the Fermi level according to theoretical calculations [18].

The experimental evidence for such a TRS breaking WSM state of $Co_3Sn_2S_2$ was first revealed in magnetotransport measurements [16,17]. This material exhibits not only negative longitudinal magnetoresistance (LMR) [16], but also large intrinsic anomalous Hall effect (AHE) [16,17] and large anomalous Nernst effect (ANE) [80,81,82]. Negative LMR is the manifestation of the chiral anomaly arising from the charge pumping between paired Weyl nodes with opposite chirality under parallel electric and magnetic fields. The intrinsic origin of AHE in $Co_3Sn_2S_2$ is evidenced by the observations that its anomalous Hall conductivity $\sigma_{yx}^A$ is nearly independent of longitudinal conductivity $\sigma_{xx}$ below 90K [16] and linearly increases with magnetization [17]. Besides large $\sigma_{yx}^A$ (~1130 $\Omega^{-1}.cm^{-1}$ at ~90K), the anomalous Hall angle ($\theta_H = \sigma_{yx}^A/\sigma_{xx}$) of $Co_3Sn_2S_2$ was also found to be large, ~ 20% at 90K, about one order of magnitude larger than those of typical magnetic systems [83]. As shown in Fig. 3b, such large values of $\sigma_{yx}^A$ and $\theta_H$ can be attributed to large net Berry curvature of occupied states [16]. The steep decrease of $\sigma_{yx}^A$ above 90 K is due to the fact that the out-of-plane FM phase coexists with the in-plane AFM phase and the volume fraction of the FM phase decreases with increasing temperature [79]. Furthermore, systematic studies on the ANE of $Co_3Sn_2S_2$ by Ding et al. [80] show that the anomalous Nernst response $S_{xy}^A$ (the ratio of transverse electric field to the longitudinal temperature gradient) is inversely proportional to the carrier mobility $\mu$, contrasted with the ordinary Nernst response $S_{xy}^0$, which is $\propto \mu$. This indicates that anomalous transverse thermoelectricity $\alpha_{xy}^A$ in $Co_3Sn_2S_2$ is determined by the Berry curvature, rather than the mean free path [80].

The hallmark of the electronic structure of a WSM phase is the surface Fermi arcs (SFAs), which connect the projected Weyl nodes with opposite chirality on the surface Brillouin-zone. Such an expected feature for $Co_3Sn_2S_2$ has recently been demonstrated by the angle-resolved photoemission spectroscopy (ARPES) [18] and scanning tunneling spectroscopy (STS) experiments [19]. As shown in Fig. 3c and 3d(i), the surface Fermi arcs are comprised of three line-segments which connect the projected Weyl points (WP+ and WP-) near M. These three line-segments form a triangle-shaped surface Fermi surface, which is clearly visualized in the ARPES [Fig. 3d(ii)] and STS experiments [18,19]. Moreover, the STS experiments also show the surface Fermi-arc contour and Weyl node connectivity is termination dependent [19]. By means of in situ electron doping, the ARPES experiments also detected bulk Weyl nodes (Fig. 3e) [18].

**1b. Heusler alloy FM WSM $Co_2MnGa$ and $Co_2MnAl$:** Recent theoretical work predicted that Co-based Heusler compound $Co_2XZ$ (X=V, Zr, Nb, Ti, Mn, Hf; Z =Si, Ge, Sn, Ga and Al) can host unique FM WSM phases [84,85,86]. First, its Weyl states can have the least number of Weyl nodes (two), which can make the interpretation of spectroscopic and transport properties much easier. Second, the Weyl node separation in momentum space is large, giving rise to large anomalous Hall effect, which is of great use for applications. Third, the Weyl node location in momentum space can be manipulated by controlling the magnetization direction[84,85]. These characteristics make $Co_2XZ$ a promising platform for exploring novel magnetic Weyl physics and potential applications.

Although Co$_2$XZ allows for many different element combinations, experimental studies on their possible exotic properties induced by the expected WSM states are sparse, which is possibly due to the difficulty of the single crystal growth of this family of materials. Co$_2$MnGa is the first confirmed member that have distinct properties associated with the FM WSM state [20,76]. This material is a room temperature ferromagnet with the Curie temperature of 690 K and possesses a cubic structure with the space group of Fm-3m (Fig.4a) [20]. It shows a giant AHE, with its $\sigma_{yx}^A$ being as large as 2000 Ω$^{-1}$ cm$^{-1}$ at low temperature (Fig.4b) [20]. Furthermore, Co$_2$MnGa was also found to show a giant ANE [20,76]. The Nernst signal S$_{yx}$ increases with elevating temperature, reaching a record high value of $S_{yx} \approx 6$ μV K$^{-1}$ at room temperature and approaching 8 μV K$^{-1}$ at 400 K [20,76], which is more than one order of magnitude larger than the typical values known for the ANE in other magnetic conductors. These results, together with the unsaturated positive longitudinal magnetoconductance (i.e. chiral anomaly) [20], provide transport evidence for Weyl fermions in Co$_2$MnGa.

Recent ARPES studies on Co$_2$MnGa [21] unveil its characteristics of Weyl state. The combination of mirror symmetries and FM ordering of this material leads to 3D Weyl nodal lines with 2-fold degeneracy, which form Hopflike links and nodal chains [21,87]. These nodal lines are protected by mirror symmetries and give rise to drumhead surface states. Both Weyl nodal lines and drumhead surface states have been visualized in the ARPES experiments [21]. The top panel in Fig. 4c shows an ARPES constant energy surface [21], from which the projection of Weyl nodal lines on the $k_a$-$k_b$ plane can be seen clearly. The distribution of calculated Berry curvature on the $k_a$-$k_b$ plane (bottom panel of Fig. 4c) matches well the shape of Weyl nodal line projection [21], indicating that the Berry curvature of Co$_2$MnGa predominantly stems from Weyl nodal lines. The $\sigma_{yx}^A$ calculated from the Berry curvature is indeed consistent with the experimental value [21].

Like Co$_2$MnGa, the L2$_1$ structural phase of Co$_2$MnAl is also predicted to be a FM WSM candidate [86] and early Berry curvature calculations suggest it has the largest AHE among the Co$_2$XZ Heusler alloys [88]. The recent success of single crystals growth of this material has enabled further experimental studies on this material. Li et al. [22] indeed observed a tunable giant AHE in Co$_2$MnAl single crystals. Its $\sigma_{yx}^A$ is as large as 1300 $\Omega^{-1}$cm$^{-1}$ at room temperature; more noticeably, its room temperature anomalous Hall angle reaches a record value among magnetic conductors, with tan$\theta_H$ = 0.21, which brings the promise for practical device applications. Theoretical studies have further clarified the intrinsic mechanism of such a giant AHE [22]. As shown in Fig. 4e, the two lowest conduction bands and the two highest valence bands cross along four Weyl nodal rings without SOC. These four nodal rings at the $k_{x,y,z} = 0$ planes are protected by the mirror symmetries and interconnected, forming Hopflike links. When SOC is considered, magnetic moments are coupled to the lattice, thus reducing the symmetry and gapping the nodal rings. The gapped nodal rings generate large Berry curvature and thus give rise to the huge anomalous Hall conductivity. For instance, when the magnetization is oriented along the [001] direction, the nodal rings on the $k_z$=0 plane do not open gaps due to the preserved mirror symmetry, while the nodal rings on the $k_{x,y} = 0$ planes are gapped. Berry curvature calculations showed the largest nodal ring (ring#3 in Fig. 4e) make dominant contributions to the Berry curvature and the resulting large $\sigma_{yx}^A$ (~1400 $\Omega^{-1}$.cm$^{-1}$) [22].

Another important property of Co$_2$MnAl is that its band topology and resulting AHE can be controlled by the rotation of magnetization axis. This is because that the mirror symmetries depend on the magnetization orientation. For example, for $M$//[001], the nodal rings on the $k_z$=0 plane are gapless, but gapped on the $k_{x,y}=0$ planes; however, if $M//[111]$, all nodal rings are gapped. Therefore, the rotation of magnetization leads to a cosθ-like angular dependence in $\sigma_{yx}^A$, which is indeed observed in experiments (Fig. 4f) [22]. Since $Co_2MnAl$ is a soft ferromagnet, the rotation of magnetization can be driven by a weak magnetic field. As such, this material offers an ideal platform to explore band topology tuning by magnetization.

2) **Antiferromagnetic WSMs**

   **2a. WSM state in Chiral antiferromagnet $Mn_3Sn$:** Besides FM materials hosting Weyl fermions as summarized above, recent works also revealed that Weyl fermions can also exist in antiferromagnetic materials. $Mn_3Sn$ is a recently established, remarkable example [23]. This material is a hexagonal antiferromagnet and exhibits noncollinear spin ordering with $T_N \approx 420$ K [23]. Although this material shows a very small magnetization ∼0.002 $\mu_B$/Mn, it exhibits large AHE [89] and ANE [90]. In addition, it also exhibits positive magnetoconductance under parallel electric and magnetic fields [23]. These distinct transport properties are associated its Weyl state which have been confirmed by both theory calculations and ARPES measurements [23]. In this material, the electron–hole band crossings form a nodal ring surrounding $K$ points without SOC. When SOC is considered, the TRS breaking lifts the spin degeneracy and leads to band crossing and the formation of Weyl nodes at different energies. Since its Weyl nodes appear at points where electron and hole pockets intersect, the resulting Weyl cones are strongly tilted, which is a typical nature of a type-II Weyl semimetal  Isostructural compound $Mn_3Ge$ is also found to host a similar WSM state [77,91]. In addition, another layered, AFM compound $YbMnBi_2$ with square lattice was also reported to have a TRS breaking, type-II Weyl state [25].

### 2b. Magnetic-field-induced WSM in AFM Half-Heusler alloys.

Another route to generate TRS breaking WSMs is to use external magnetic field to generate Weyl nodes in Dirac semimetals or zero gap semiconductors. For instance, bulk Dirac cones in $Na_3Bi$ have been found to evolve to Weyl cones under magnetic field [92]. The first example of Weyl nodes generated by external magnetic fields in zero gap semiconductors is GdPtBi [24, 93,94], which is a half Heusler compound and possesses a cubic structure consisting of interpenetrating face centered cubic lattices and exhibits antiferromagnetic ordering with $T_N =$ 9.2 K [24]. The hallmarks of Weyl state in transport, including negative LMR caused by chiral anomaly [24], large intrinsic AHE [93] and planar Hall effect (PHE) [95], have been demonstrated in this material. Such a magnetic field-driven Weyl state is believed to originate either from the Zeeman splitting by the external magnetic field [24] or from the exchange splitting of the conduction bands [94]. The finding of field-induced Weyl state in GdPtBi has inspired studies on other isostructural half Heusler compounds like NdPtBi [94] and TbPtBi [96,97]. TbPtBi was found to show an exceptionally large AHE with the anomalous Hall angle of 0.68-0.76 [96] ( about a few times larger than that in GdPtBi [93]), though its other transport signatures of Weyl state (e.g. PHE) are not significant. The first-principle electronic structure and the associated anomalous Hall conductivity calculations show that the exceptionally large AHE in TbPtBi does not originate from the Weyl points but that it is driven by the large net Berry curvature produced by the anticrossing of spin-split bands near the Fermi level [96].

### 3. Outlook

From the above overview, it can be seen clearly that the interplay between band topology and magnetic states can create unique topological quantum states which are potentially useful for technology applications. QAHI is the most promising example. However, device applications require such a state to be realized at room temperature. Although theoretical studies show this is possible, efforts are needed to discover more promising candidate materials which combine band topology with room temperature magnetism. Among the current magnetic TIs, the magnetic transitions all occur at low temperatures (below 50 K). TIs with room temperature magnetism is highly desired. Materials design by theory and computations could play a key role in this regard. Since FM WSMs can also evolve into QAHI in the 2D limit, discovering room temperature ideal FM WSMs may be another route to realize high temperature QAHI. An ideal WSM generally refers to a Weyl state with all Weyl nodes being symmetry related and at or close to the chemical potential, without interfered with by any other bands; current magnetic WSMs are not ideal. Furthermore, magnetic topological materials offer unique opportunities to explore the topological-electronic-state's tunability by magnetism and new fundamental physics of topological fermions.

**Acknowledgement:**

Z.Q.M. acknowledges the support from the US National Science Foundation (NSF) under grants DMR 1917579, 1832031 and the Penn State 2D Crystal Consortium-Materials Innovation Platform (2DCC-MIP) funded by the NSF (cooperative agreement DMR-1539916).

**Data Availability Statements**：**The data that support the findings of this study are available from the corresponding author upon reasonable request.**

**Captions**

Figure 1. (a) Crystal and magnetic structure of MnBi$_2$Te$_4$ [29]. (b) Spin-resolved electronic structure of the MnBi$_2$Te$_4$ (0001) surface.[31] (c) The bulk band structure of MnBi$_2$Te$_4$ in the FM phase, which shows a pair of Weyl nodal crossing at W/W′ along the Γ-Z direction [30]. (d) Band dispersion of MnBi$_2$Te$_4$ measured on the (0001) surface [31]. (e) A gapless surface Dirac

cone measured on MnBi$_2$Te$_4$ [39]. (f) The maximal anomalous Hall angle of the FM phase vs. carrier concentration for Mn(Bi$_{1-x}$Sb$_x$)$_2$Te$_4$ [50].

Figure 2. (a) Magnetic field dependences of Hall resistance $R_{yx}$ and longitudinal resistance $R_{xx}$ measured on a flake with five SLs at $T = 1.4$ K. $R_{yx}$ reaches $0.97h/e^2$ at $\mu_0 H = 0$ T and rises to $0.998h/e^2$ above $\mu_0 H \sim 2.5$ T [45]. (b) The gate voltage ($V_g$) dependence of longitudinal resistivity $\rho_{xx}$ and the derivative of Hall resistivity $\rho_{yx}$ with respect to magnetic field measured at $T = 1.6$ K around zero magnetic field, measured on a 6 SLs device.[46]. (c) The gate voltage dependence of $\rho_{xx}$ and $\rho_{yx}$ at 1.6 K and 9 T, from which typical characteristics of a Chern insulator can be seen. The inset in (b) and (c) schematically illustrates the FM order and electronic structures of the Axion and Chern insulators [46]. (d) Hall resistance $R_{yx}$ as a function of magnetic field at various temperatures probed in a 10-SL MnBi$_2$Te$_4$ device. The $R_{yx}$ plateau reaches $0.97h/2e^2$ at 13 K [47]. (e) Gate voltage dependence of $R_{xx}$ and $R_{yx}$ at 2 K and 15 T in a 10-SL device. The inset schematically illustrates the FM order and electronic structure of the $C$=2 Chern insulator [47].

Figure 3. (a) Crystal structure of Co$_3$Sn$_2$S$_2$. The cobalt atoms form a ferromagnetic kagome lattice with a $C_{3z}$-rotation [16]. (b) Temperature dependences of the anomalous Hall conductivity ($\sigma_H^A$), the longitudinal conductivity ($\sigma$) and the anomalous Hall angle ($\sigma_H^A/\sigma$) at zero magnetic field [16]. (c) Schematic of the bulk and surface Brillouin zones along the (001) surface of Co$_3$Sn$_2$S$_2$, which display three pairs of Weyl nodes connected by surface Fermi arcs (SFA, marked by yellow line segments) [18]. (d) Comparison between the calculated Fermi surface of both bulk and surface states (i) and the experimentally measured Fermi surfaces (ii). The magenta and green dots in (i) represent the Weyl points with opposite chirality [18]. (e) Linear band crossing at a Weyl point probed by the ARPES measurements at 10K [18].

Figure 4. (a) L2$_1$ ordered cubic full Heusler structure of Co$_2$MnGa [20]. (b) Temperature dependence of the Hall conductivity -$\sigma_{yx}$ for Co$_2$MnGa [20]. (c) The bottom panel shows the $z$ component of the Berry curvature of occupied states and the top panel presents the ARPES constant energy surface at the corresponding $E_B$ [21]. (d) Nodal rings and the first Brillouin zone of Co$_2$MnAl. Without SOC, there are nodal rings on mirror planes [22]. (e) There are four nodal rings centered at the Z point of the FCC Brillouin zone for Co$_2$MnAl [22]. (f) The anomalous Hall conductivity $\sigma_{yx}^A$ of Co$_2$MnAl as a function of the magnetization orientation angle. The experimental and theoretical results are represented by red and black circles, respectively [22].

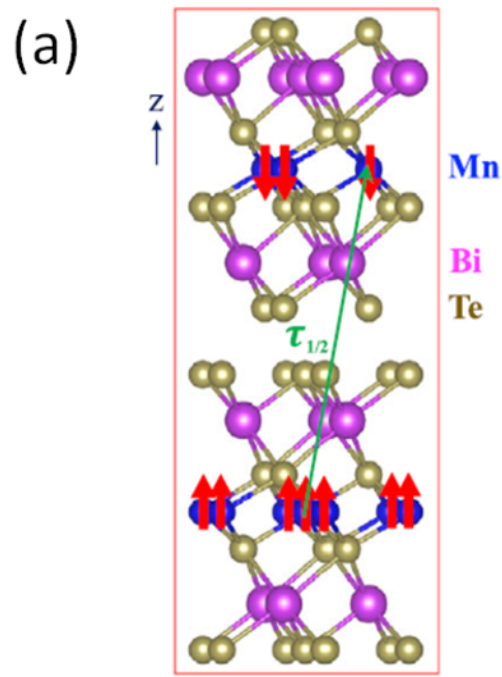 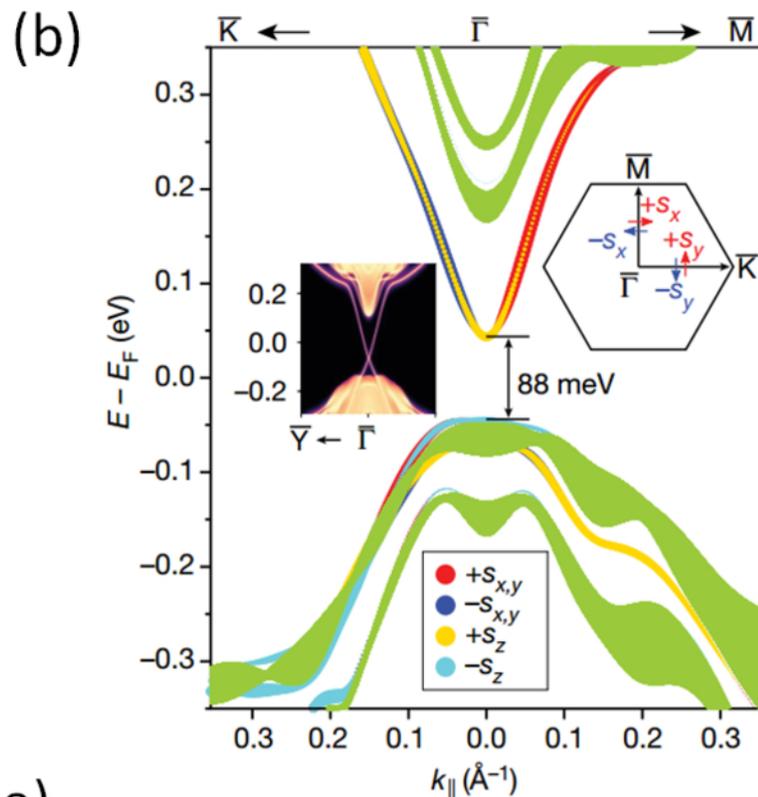 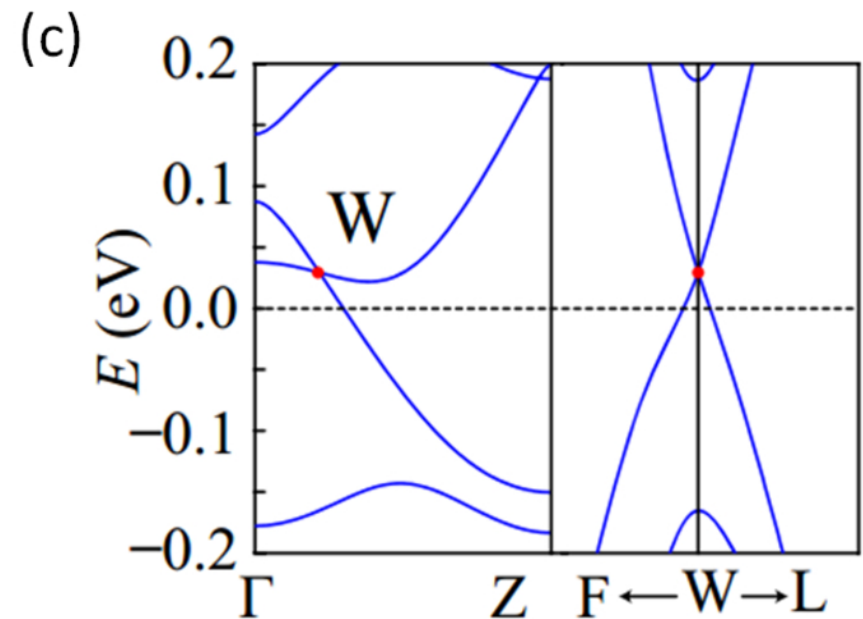
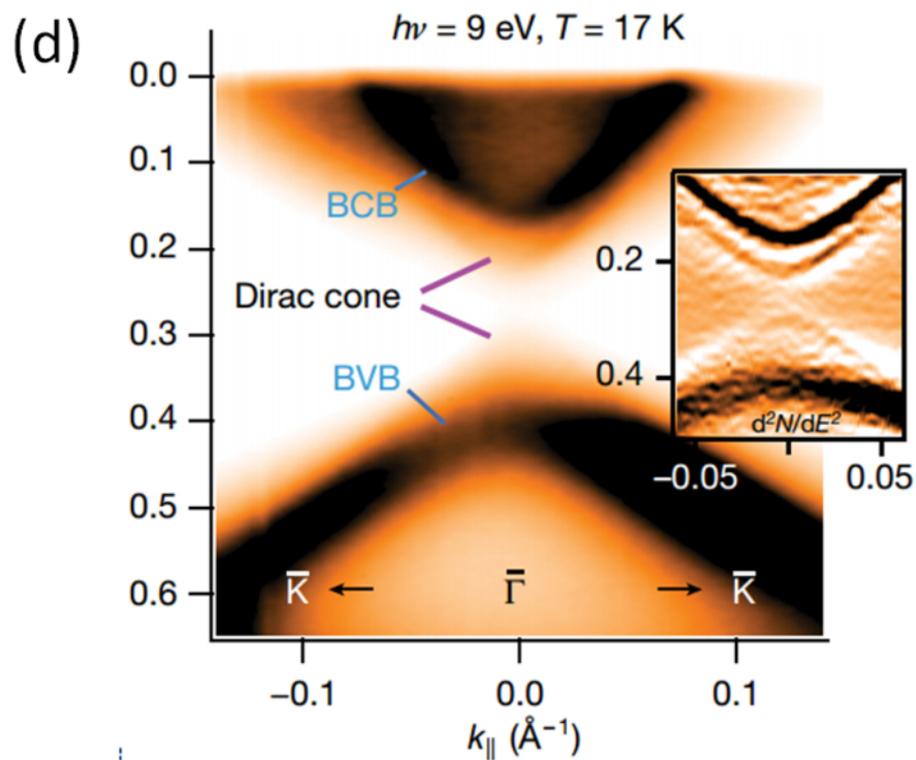 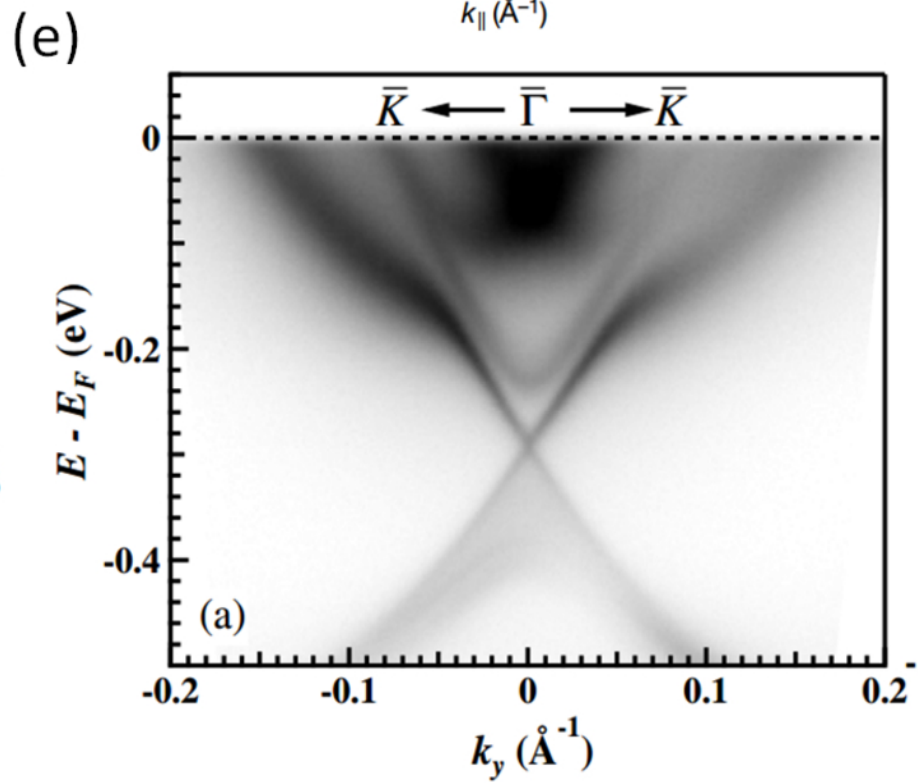 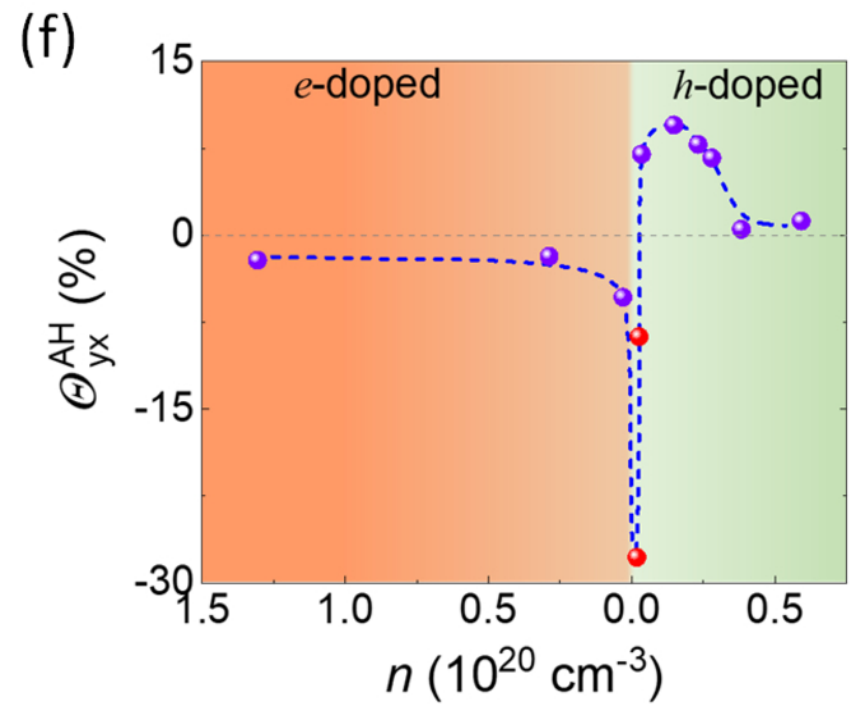

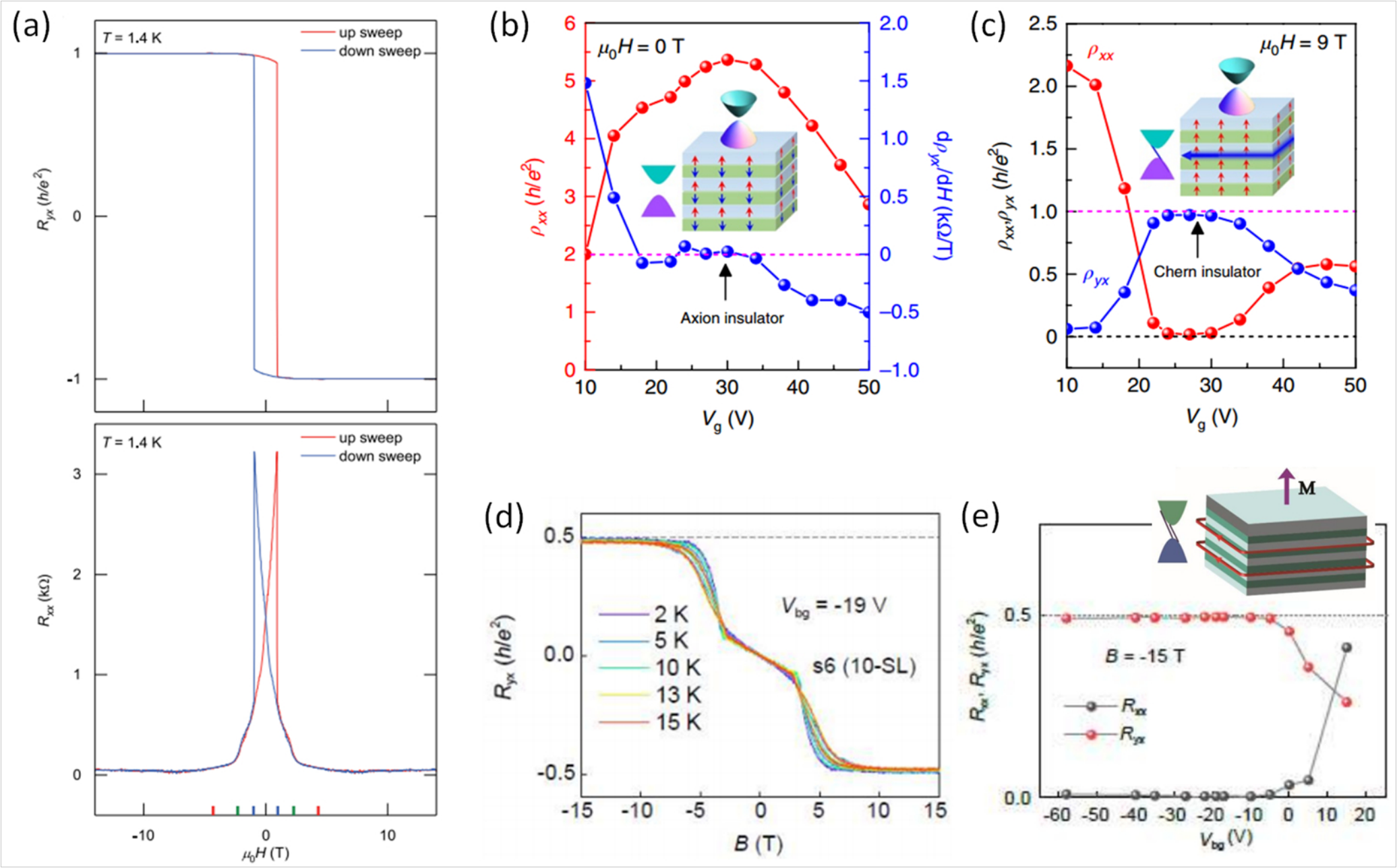

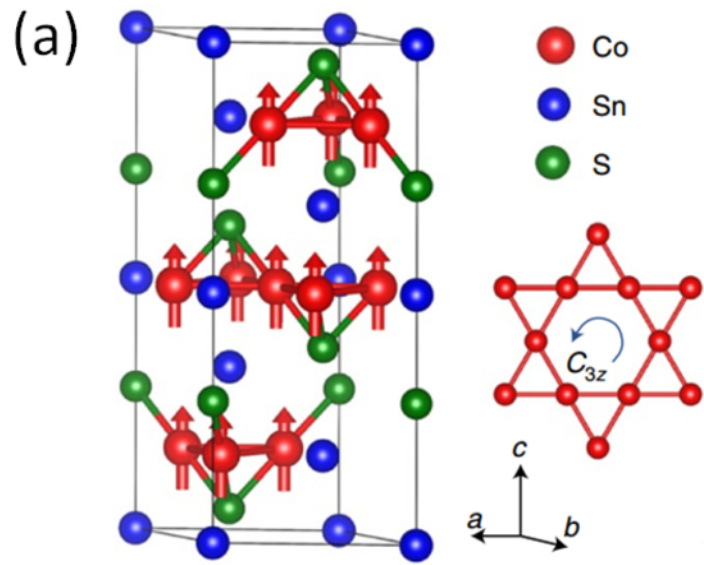
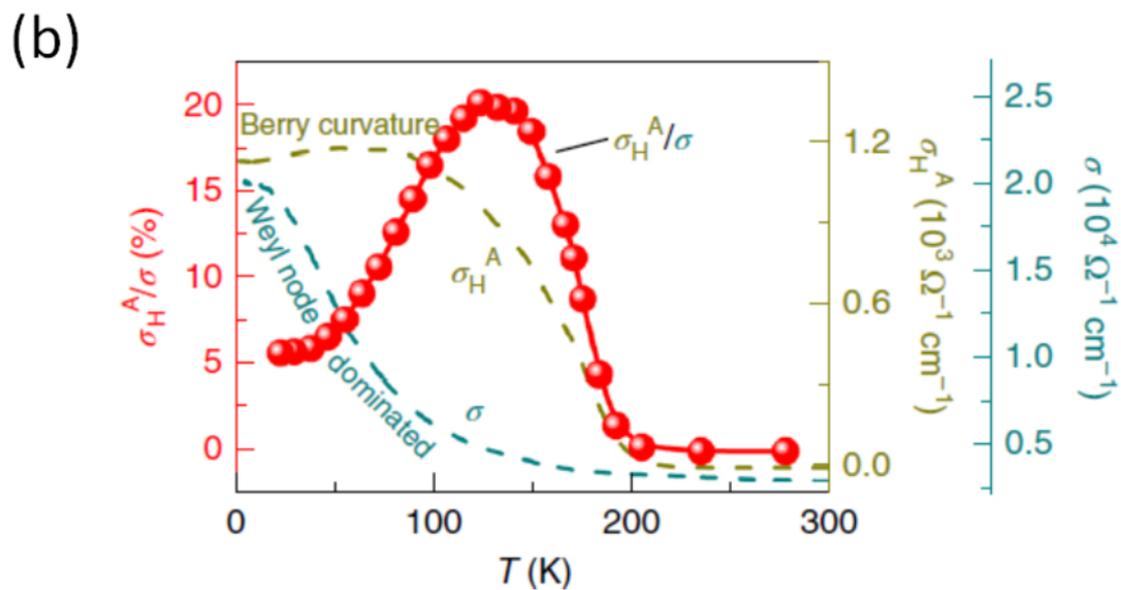
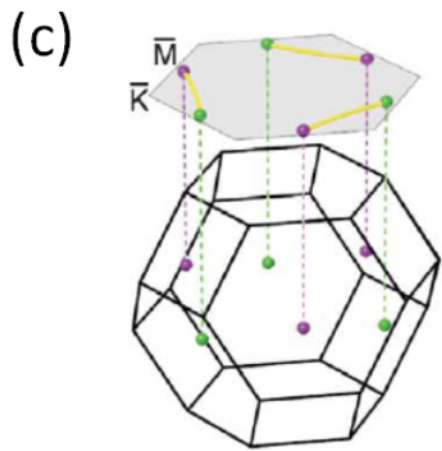
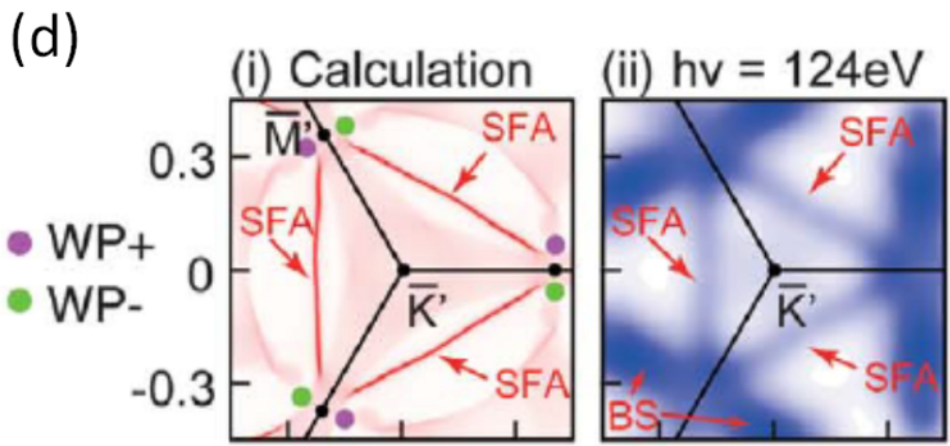
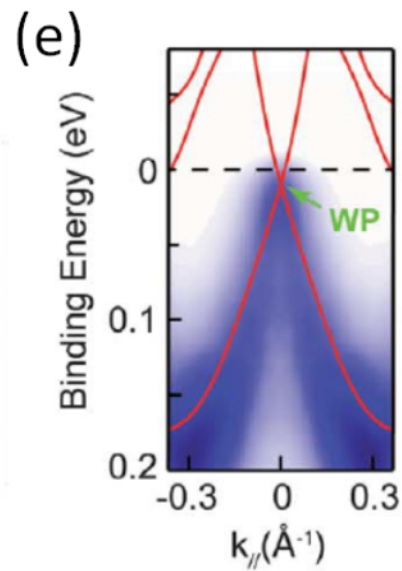

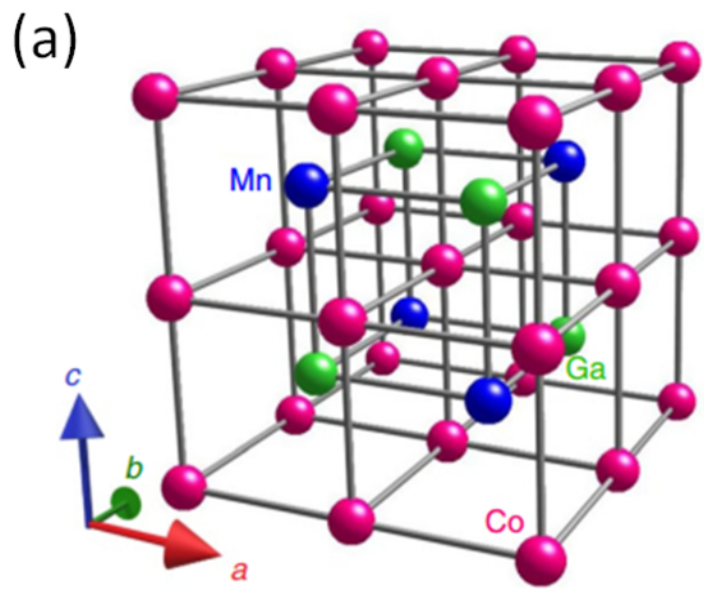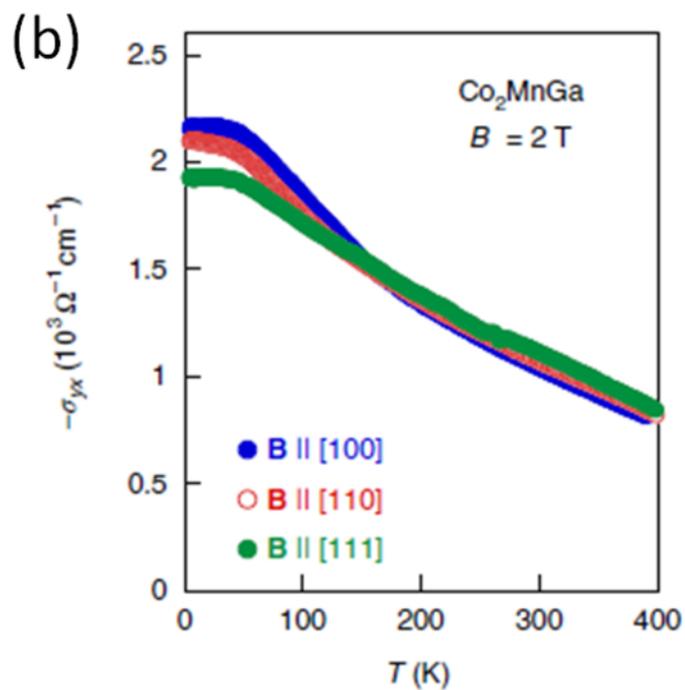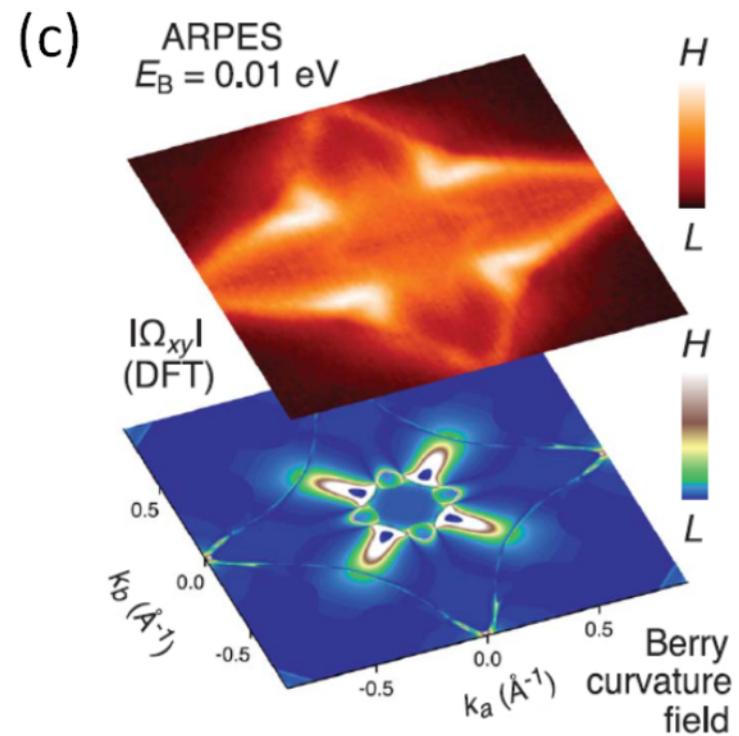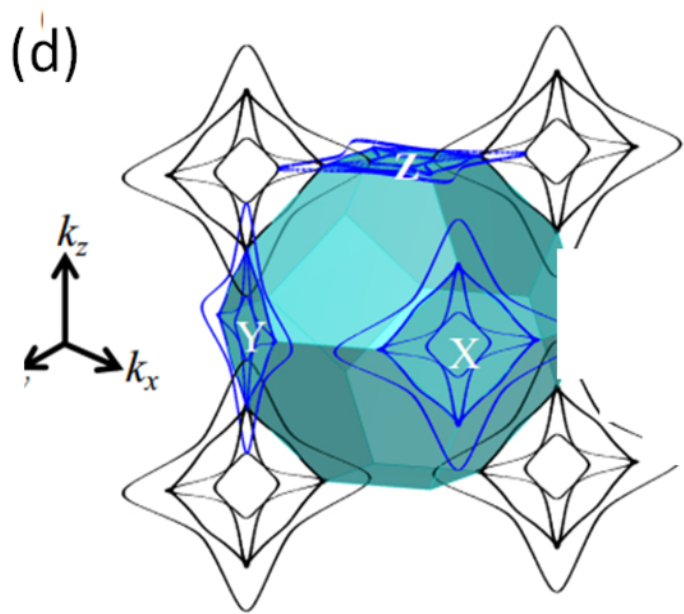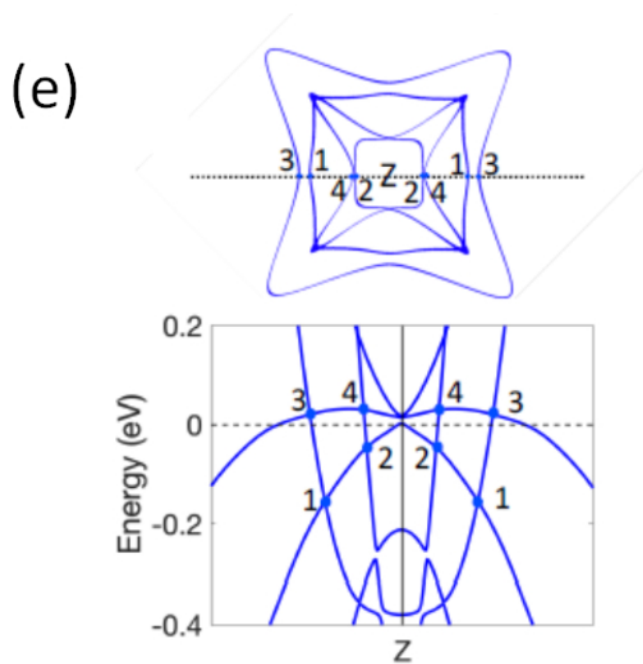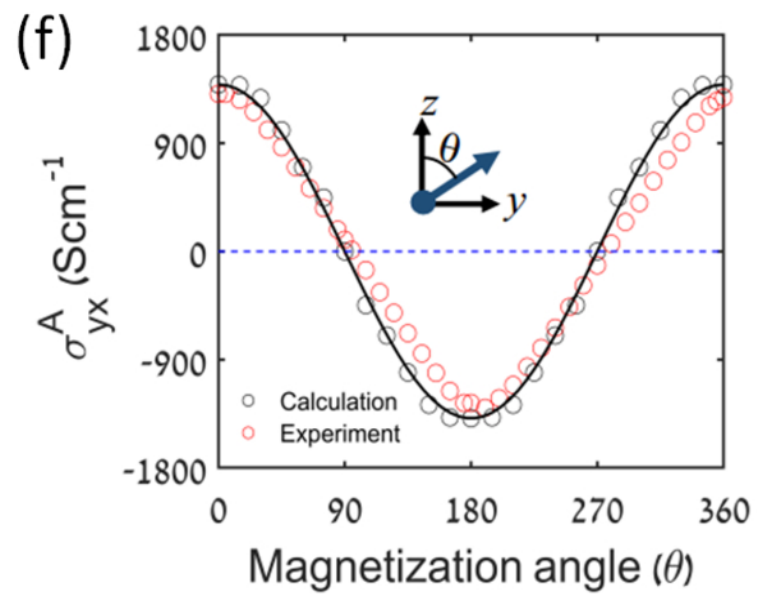